\documentstyle[11pt]{article}
\date{}
\newcommand{\eeq}{\end{eqnarray}}
\newcommand{\beq}{\begin{eqnarray}}
\newcommand{\bD}{{\bf D}}

\newcommand{\bB}{{\bf B}}

\newcommand{\cH}{{\cal H}}

\newcommand{\p}{{\partial}}

\newcommand{\bY}{{\bf Y}}
\newcommand{\bX}{{\bf X}}

\def\con{{}_{\_\rule{-1pt}{0pt}\_}
\rule{-2pt}{0pt}\raise1.5pt\hbox{$\mid$}\hspace{2pt}}
\title{\bf Quantization conditions from the group theory}
\author{Dariusz Chru\'sci\'nski\footnotemark \\
 Institute of Physics, Nicholas Copernicus University\\
 ul. Grudzi\c{a}dzka 5/7, 87-100 Toru\'n, Poland}

\begin{document}
\def\thefootnote{\relax}\footnotetext{$^*$E-mail:
darch@phys.uni.torun.pl}

\maketitle

\begin{abstract}

We show that the requirement of the relativistic invariance for any
self-interacting, abelian $p$-form theory uniquely determines the form of
the corresponding quantization condition.

\vspace{.5cm}



\end{abstract}

The old idea of electric-magnetic duality  has played in
recent years very prominent role in quantum field theory and string theory
(see e.g. \cite{Olive}).
One of its most
fascinating implication  is the celebrated
Dirac quantization condition \cite{Dirac}
\beq  \label{DQC}
eg=nh\ ,
\eeq
with integer $n$ ($h$ is the Planck constant).
This idea may be in a straightforward  manner generalized to
self-interacting $p$-form electrodynamics in $D$ dimensions
\cite{Savit}--\cite{p-form}.
Then the condition (\ref{DQC}) still holds with $e$ and $g$ being the
electric and magnetic charges of the elementary electric $(p-1)$--brane
and magnetic $(D-p-3)$--brane.

When $D=2p+2$, $(p-1)$--brane dyons may exist. In $D=4$ (i.e. for  $p=1$)
it was shown by Zwanziger and Schwinger \cite{ZS} that the Dirac condition
(\ref{DQC}) should be replaced by
\begin{equation}  \label{QC1}
e_1 g_2 -  e_2 g_1 = nh\ .
\end{equation}
Now, for $p>1$, the quantization
condition for $(p-1)$--brane dyons crucially depend upon the parity of $p$
\cite{Deser-2}. When $p$ is odd it is simply a generalization of
(\ref{QC1}) with $e$ and $g$ being the electric and magnetic charges of
$(p-1)$--brane dyon. But when $p$ is even one has to replace (\ref{DQC})
by
\begin{equation}  \label{QC2}
e_1 g_2 +  e_2 g_1 = nh\ .
\end{equation}
In \cite{Deser-2} both conditions were derived by introducing the
 Dirac $p$-branes (the generalization of  Dirac strings \cite{Dirac}) and
taking into account the multiple
connectedness of the configuration space of Dirac $p$-branes and
$(p-1)$--brane dyons.
It should be stressed that both conditions
are valid for any self-interacting (not only Maxwell one), gauge-invariant
electrodynamics irrespective of its duality invariance.

In the present paper we show that knowing the Dirac condition
(\ref{DQC}), both conditions (\ref{QC1}) and (\ref{QC2}) follow
immediately from the group theory. Our argument goes as follows: note
first, that the Dirac condition (\ref{DQC}) may be generalized only either
to (\ref{QC1}) or (\ref{QC2}). There is no other possibility.   The main
difference between (\ref{QC1}) and (\ref{QC2}) lies in the corresponding
symmetry groups.  Note, that both conditions are invariant under the
following scaling transformations:
\beq
e &\rightarrow&  \lambda\, e\nonumber \ ,\\
g &\rightarrow&  \frac{1}{\lambda}\, g \ , \label{I}
\eeq
with $\lambda \neq 0$. But condition (\ref{QC1}), contrary to (\ref{QC2}),
 is additionally invariant under $SO(2)$ orthogonal rotations
\beq
e &\rightarrow&  e\, \cos\alpha - g\, \sin\alpha\nonumber \ ,\\  \label{II}
g &\rightarrow&  e\, \sin\alpha + g\, \cos\alpha \ ,
\eeq
and $SO(1,1)$ hyperbolic rotations
\beq
e &\rightarrow&  e\, \cosh\alpha + g\, \sinh\alpha\nonumber \ ,\\
g &\rightarrow&  e\, \sinh\alpha + g\, \cosh\alpha \ . \label{III}
\eeq
Note, that transformations
(\ref{I})--(\ref{III}) generate $SO(2,1)$ group, i.e. they
realize the $SO(2,1)$ group as linear transformations in the 2-dimensional
 space parameterized by $(e,g)$. Now, condition (\ref{QC2}) is neither
invariant under (\ref{II}) nor under (\ref{III}), but it is  invariant
under the discrete $Z_2$ transformation
\beq  \label{Z2}
e\rightarrow g\ ,\ \ \ \ \ \ g\rightarrow e\ .
\eeq
Obviously, condition (\ref{QC1}) is not $Z_2$ invariant. Since (\ref{I})
defines the $SO(1,1)$ group (subgroup of $S(2,1)$), therefore, condition
(\ref{QC1}) is $SO(2,1)$ invariant, whereas condition (\ref{QC2}) is
$SO(1,1) \times Z_2$ invariant.  Note, that (\ref{II}) are nothing but the
duality rotations between electric and magnetic charges.

Now, it turns out that the same symmetry groups are encoded into the canonical
structure of any gauge-invariant electrodynamics. Consider first $p=1$ theory
in $D=4$.
It is easy to see that the
canonical Poisson bracket
\begin{equation}  \label{P1}
\{ D^i(x), B^j(y) \} = \epsilon^{ikj} \p_k \delta^{(3)}(x-y)\ ,
\end{equation}
is invariant under  the same group of transformations
(\ref{I})--(\ref{II}) with $(e,g)$ replaced by $(\bD,\bB)$. Physically, the
$SO(2,1)$ group is implied be the relativistic invariance. In the Hamiltonian
framework this invariance is equivalent to the symmetry of the corresponding
energy-momentum tensor. Now, any Hamiltonian is a functional of the following
three scalar functions built out of $\bD$ and $\bB$ \cite{IBB}
(see \cite{STRONG} for a $p$-form generalization):
\beq
\alpha &=& \frac 12 (\bD^2 + \bB^2)\ ,\\
\beta &=& \frac 12 (\bD^2 - \bB^2)\ ,\\
\gamma &=& \bD\bB\ .
\eeq
The symmetry of $T^{\mu\nu}$ implies the following equation for the
Hamiltonian $\cH$:
\begin{equation}   \label{H1}
(\p_\alpha \cH)^2 - (\p_\beta \cH)^2 - (\p_\gamma \cH)^2 = 1\ ,
\end{equation}
which displays exactly the $SO(2,1)$ symmetry. Therefore, transformations
(\ref{I})--(\ref{II}) with $(e,g)$ replaced by $(\bD,\bB)$ are nothing but a
realization in the space of fields $(\bD, \bB)$ of this {\it fundamental}
symmetry.
Summarizing:  the invariance group of the standard
Dirac-Zwanziger-Schwinger condition (\ref{QC1}) is implied by the
relativistic invariance of the underlying (possibly nonlinear)
electrodynamics.

Now, we demand that the same property holds for any $p$-form theory, i.e.
that the symmetry group of the corresponding quantization condition is
 implied by the relativistic invariance of the corresponding
$p$-form theory. When $p$ is odd any Hamiltonian is built out of $\alpha$,
$\beta$ and $\gamma$, where now $\bX\bY =
\frac{1}{p!}\, X^{i_1...i_p}Y_{i_1...i_p}$ for any $p$-forms $X$ and $Y$.
Therefore, all arguments are the same as for $p=1$ and the corresponding
quantization condition is given by (\ref{QC1}). When $p$ is even the
situation is different. Now $\bD\bB=0$ and, therefore, any Hamiltonian
depends only upon $\alpha$ and $\beta$. The symmetry of $T^{\mu\nu}$ gives
rise to
\begin{equation}   \label{H2}
(\p_\alpha \cH)^2 - (\p_\beta \cH)^2  = 1\ ,
\end{equation}
which, contrary to (\ref{H1}),  displays only  the $SO(1,1)$ symmetry.
This group is realized in the space of fields ($p$-forms) $(D,B)$ as the
group of canonical transformations. It is easy to see that the canonical
Poisson bracket   (a $p$-form generalization of (\ref{P1}))
\begin{equation}  \label{Pp}
\{ D^{i_1...i_p}(x), B^{j_1...j_p}(y) \} = \epsilon^{i_1...i_p k
j_1...j_p} \p_k \delta^{(2p+1)}(x-y)\ ,
\end{equation}
 is (for even $p$) no longer invariant under the full $SO(2,1)$ group
but  only under the $SO(1,1)$ subgroup generated by  (\ref{I}) (with
$(e,g)$
replaced by $(D,B)$). Moreover, since for any $p$-form theory
\beq
T^{0k} &=& \frac{1}{p!} (-1)^{p+1} \, \epsilon^{ki_1...i_pj_1...j_p}\,
D_{i_1...i_p}B_{j_1...j_p}\ ,\\
T^{k0} &=& \frac{1}{p!} (-1)^{p+1} \, \epsilon^{ki_1...i_pj_1...j_p}\,
E_{i_1...i_p}H_{j_1...j_p}\ ,
\eeq
therefore, for even $p$ the condition $T^{0k} = T^{k0}$ is obviously invariant
under the discrete $Z_2$ transformation:
\begin{equation}   \label{D-B}
D\rightarrow B\ ,\ \ \ \ \ \ B\rightarrow D\ .
\end{equation}
Note, that (\ref{Pp}) is also invariant under (\ref{D-B}), i.e.
(\ref{D-B})  defines a canonical symmetry.

Summarizing: the relativistic invariance of any $p$-form
theory defines the $SO(2,1)$ and $SO(1,1)\times Z_2$ groups of canonical
symmetries for odd and even $p$ respectively, i.e. both canonical
structure and quantization condition have the same symmetry properties.
Therefore, the structure of the corresponding quantization condition
((\ref{QC1}) or (\ref{QC2})) is uniquely determined by the requirement of
the relativistic invariance of the underlying $p$-form theory.

\vspace{.5cm}

The author thanks Peter Orland for pointing out Refs. \cite{Savit} and
\cite{Orland} where the quantization condition for a $p$-form field was
introduced before well known papers \cite{Nepomechie} and \cite{p-form}.
This work was partially supported by the KBN Grant No. 2 P03A
047 15.


\begin{thebibliography}{99}



\bibitem{Olive} D. Olive, Nucl. Phys. (Proc. Supl. ) {\bf B 58} (1997) 43.

\bibitem{Dirac} P. A. M. Dirac, Proc. Roy. Soc. {\bf A 133} (1931) 60;
Phys. Rev. {\bf 74} (1948) 817.

\bibitem{Savit} R. Savit, Phys. Rev. Lett. {\bf 39} (1977) 55.

\bibitem{Orland} P. Orland, Nucl. Phys. {\bf B 205} (1982) 107.

\bibitem{Nepomechie} R. Nepomechie, Phys. Rev. {\bf D 31} (1985) 1921.

\bibitem{p-form}
 C. Teitelboim, Phys. Lett. {\bf B 167} (1986) 63, 69;
 M. Henneaux, C. Teitelboim,  Foundations of Physics
{\bf 16} (1986) 593.


\bibitem{ZS} D. Zwanziger, Phys. Rev. {\bf 176} (1968) 1489;
\par\noindent

J. Schwinger,
Phys. Rev. {\bf 173} (1968) 1535; Science {\bf 165} (1969) 757.



\bibitem{Deser-2} S. Deser, A. Gomberoff, M. Henneaux and C. Teitelboim,
 Nucl. Phys. {\bf B 520} (1998) 179.

\bibitem{IBB} I. Bia{\l}ynicki-Birula, {\it Nonlinear Electrodynamics: Variations
on a Theme by Born and Infeld}, in {\it Quantization Theory of Particles and
Fields}, (ed. B. Jancewicz and J. Lukierski, World-Scientific, 1983).


\bibitem{STRONG} D. Chru\'sci\'nski, {\it Strong field limit of the
Born-Infeld p-form electrodynamics}, hep-th/0005215 (to appear in Phys. Rev.
D).

\end{thebibliography}
\end{document}